\documentclass[a4paper,11pt]{article}
\pdfoutput=1	
\usepackage{jheppub}
\usepackage{graphicx,wasysym}
\usepackage{slashed}
\usepackage{soul,xcolor}
\newcommand{\N}{\ensuremath{\mathbbm{N}}}

\renewcommand{\a}{\alpha}

\renewcommand{\d}{\delta}

\newcommand{\p}{\partial}
\newcommand{\lb}{\bar{\lambda}}
\newcommand{\la}{\lambda}
\newcommand{\pb}{\bar{\psi}}
\newcommand{\ps}{\psi}

\newcommand{\s}{\sigma}

\newcommand{\e}{\varepsilon}
\newcommand{\f}{\phi}

\newcommand{\be}{\begin{equation}}
\newcommand{\ee}{\end{equation}}
\newcommand{\bea}{\begin{align}}
\newcommand{\eea}{\end{align}}
\newcommand{\li}{\hspace{1mm}}

\newcommand{\rf}[1]{(\ref{#1})}

\def\E {$E_{7(7)}$}
\def\N{$\mathcal{N}$}

\usepackage{color}

\title{\boldmath Origin of Soft Limits from Nonlinear Supersymmetry in Volkov--Akulov Theory}

\author[]{Renata Kallosh,}
\author[]{Anna Karlsson}
\author[]{and Divyanshu Murli}

\affiliation[]{Stanford Institute for Theoretical Physics and Department of Physics, \\ Stanford University, Stanford, CA 94305, U.S.A.}

\emailAdd{kallosh@stanford.edu}
\emailAdd{annakarl@stanford.edu}
\emailAdd{divyansh@stanford.edu}

\abstract{We apply the background field technique, recently developed for a general class of nonlinear symmetries, at tree level, to the Volkov--Akulov theory with spontaneously broken $\mathcal{N}=1$ supersymmetry. We find that the background field expansion in terms of the free fields to the lowest order reproduces the nonlinear supersymmetry transformation rules. The double soft limit of the background field is, in agreement with the new general identities, defined by the algebra of the nonlinear symmetries.}

\keywords{Supersymmetric Effective Theories, Scattering Amplitudes, Spontaneous Symmetry Breaking}

\begin{document}
\setstcolor{red}
\frenchspacing
\maketitle
\flushbottom

\section{Introduction}
In models with linear symmetries there are well-known Ward identities constraining the physical observables, both at tree level and regarding quantum corrections, in the absence of anomalies. In the case of nonlinear symmetries, often associated with some coset space ${G/ H}$, it is known that certain soft theorems relate on-shell amplitudes with different number of legs to each other. However, until recently there was no general analogue of the Ward identities for the case of nonlinear, spontaneously broken symmetries, for example in Volkov--Akulov (VA) theory \cite{Volkov:1973ix}. Such a general analysis of nonlinear symmetries was performed at tree level in \cite{Kallosh:2016qvo}, where the relevant Ward type identities were derived through a generalization of the background field method in the abstract formalism of DeWitt \cite{DeWitt:1967ub,Kallosh:1974yh,Grisaru:1975ei}. The construction provides an initial method for analyses, with a possible, later continuation to loops. The new identities in \cite{Kallosh:2016qvo} should be applicable to any model with nonlinear symmetry\footnote{The background field method for NSM (nonlinear sigma models) was developed in \cite{Honerkamp:1971sh}. General models with single and multiple Nambu--Goldstone bosons were studied recently in \cite{Low:2014nga,Low:2015ogb}.} where the action and its nonlinear symmetries are known, and the transformation rule has a constant field-independent part as well as various powers of the fields beyond the linear dependence. Examples for which the new identities should be valid include the VA theory with fermionic, nonlinear supersymmetry \cite{Volkov:1973ix}, \N$=8$ supergravity with bosonic, nonlinear \E\, symmetry \cite{Cremmer:1979up}, and the Dirac--Born--Infeld--Volkov--Akulov (DBI--VA) model with 16 linear $+$ 16 nonlinear supersymmetries, as presented in \cite{Bergshoeff:2013pia}.

The purpose of this article is to test these new, general identities, applicable to such different examples, in the simplest setting possible: the VA model. We will study the nonlinear supersymmetry and the relation between the symmetry and the soft limits for the on-shell amplitudes in the VA theory, to lowest order in the free fields, guided by the general identities derived in \cite{Kallosh:2016qvo}.

In \cite{Akulov:1974xz}, it was shown that the S-matrix in the fermionic VA theory satisfies Adler's principle, i.e. that elements of the on-shell S matrix tend to zero when any of the four-momenta of the fermionic goldstino tend to zero. Subsequently, the VA low-energy theorems, including the single and double soft limits, were studied in \cite{deWit:1975xci}, and it was shown that the VA double soft limit is given by the supersymmetry algebra. The single and double soft limits in the case of a bosonic, nonlinear \E\, symmetry in \N$=8$ supergravity were studied in \cite{ArkaniHamed:2008gz,Kallosh:2008rr,Brodel:2009hu}. However, it was only recently realized \cite{Chen:2014xoa} that the double soft limits in those two theories are of the same nature \cite{ArkaniHamed:2008gz,Brodel:2009hu}. Namely, that the double soft limit is defined by an algebra of the spontaneously broken symmetry generators $G$ resulting in an unbroken symmetry $H$, symbolically
\be
[G, G] = H\li,
\ee
with $G$ the generators of \E\, and $H$ the generators of $SU(8)$, or, in the VA model, $G$ the fermionic super-Poincar\'e and $H$ a bosonic translation \cite{Akulov:1974xz}.

Prior to \cite{Kallosh:2016qvo}, this universality of the double soft limit was an observation on the structure of amplitudes explicitly constructed either using Feynman rules as in \cite{Chen:2014xoa}, recursion relations as in \cite{Luo:2015tat}, or the CHY scattering equations, as in \cite{He:2016vfi,Cachazo:2016njl}. In \cite{Kallosh:2016qvo}, an explanation of the universality was provided: it was shown why the algebras of these symmetries show up in the double soft limit in the solution for a background field, as a functional of the free fields.

In the context of string theory the Volkov--Akulov Lagrangian can be obtained by gauge fixing the $\kappa$-symmetric D3 brane action, which makes that model very interesting, as it is related to a fundamental way of spontaneous supersymmetry breaking in string theory. 

\section{The VA model}
The Lagrangian in the $D=4$, $\mathcal{N}=1$ VA model\footnote{The notation follows \cite{Clark:1988es}, except that we have set the goldstino decay constant $\kappa = 1$ for simplicity. In addition, we use the unconstrained field $\psi$ in the action and reserve $\la$ for the on-shell, free field, as it is often used in amplitudes. Our spinor indices will be $(a, \dot a)$ to avoid a confusion with the symmetry parameters $\xi^\alpha$ of the background field method.} is
\begin{equation}\label{VALag}
\mathcal{L} = -\frac{1}{2} \det A\li,
\end{equation}
where the vierbein $A$ is\footnote{Throughout the article, Greek letters will denote spacetime (Lorentz) indices, as in $\partial_\mu$.}
\begin{equation}
A_{\mu}{}^{\nu} \equiv \delta_{\mu}{}^{\nu} + i \psi \partial_{\mu} \sigma^{\nu} \bar \psi - i \partial_{\mu} \psi \sigma^{\nu} \bar \psi \equiv \delta_{\mu}{}^{\nu} + i \psi \overleftrightarrow{\partial_{\mu} } \sigma^{\nu} \bar \psi\li.
\end{equation}
The Lagrangian (\ref{VALag}) is invariant under the nonlinear supersymmetry variations
\begin{subequations}\label{eq.sym12}
\begin{align} \label{sym1}
\d^Q(\e, \bar{\e})\psi^{a} &= \e^{a} - i (\psi \s^{\mu} \bar{\e} - \e \sigma^{\mu} \bar \psi) \p_{\mu} \psi^{a}\li, \\ 
\d^Q(\e, \bar{\e})\bar \psi_{\dot a} &= \bar\e_{\dot a} - i (\psi \s^{\mu} \bar{\e} - \e \sigma^{\mu} \bar \psi) \p_{\mu} \bar \psi _{\dot a}\li,
 \label{sym2}\end{align}
 \end{subequations}
where $(\e^{a}$, $\bar{\e}_{\dot a})$ are infinitesimal Weyl spinor transformation parameters. The transformations form the algebra
\be
[ \d^Q(\e,\bar{\e}), \d^{Q}(\eta,\bar \eta)]= -2 i (\e\sigma^\rho \bar \eta-\eta\sigma^\rho \bar \e) \partial_\rho\li.
\label{algebra}
\ee

The variations of the action are given by 
\begin{subequations}\label{eq.eq12}
\begin{align}\label{eq1} 
\frac{\d S}{\d \psi^{a} } &= i \, T^{\mu\nu} \, (\s_\nu \p_\mu \bar \psi)_{a} + i\p_{\mu} [T^{\mu\nu} (\s_\nu \bar \psi)_{a} ]\li, \\ 
\frac{ \d S }{\d \bar \psi^{\dot a} } &= i\, T^{\mu\nu} (\p_\nu \psi \s_\mu)_{\dot a} + i\, \p_{\mu} [T^{\nu\mu} (\psi \s_\nu)_{\dot a} ]\li,
\label{eq2}\end{align}\end{subequations}
where $T^{\mu\nu}$ is the on-shell conserved Noether energy-momentum tensor associated with translation invariance
\begin{gather}\begin{aligned}
T^{\mu\nu} &= \p^{\nu} \psi^{a}\frac{\p \mathcal{L} }{\p \p_{\mu} \psi^{a} } + \frac{\p \mathcal{L} }{\p \p_{\mu} \bar \psi^{\dot a} } \p^{\nu} \bar \psi^{\dot a} - \eta^{\mu\nu}\mathcal{L} \\
&= -\frac{1}{2} \det A (A^{-1})^{\nu\mu}= -{1\over 2} \eta^{\mu\nu} + \tilde T^{\mu\nu}.
\end{aligned}\end{gather}

The Volkov--Akulov Lagrangian transforms as a total derivative under the supersymmetry transformations in \eqref{eq.sym12}
\be
\d^Q(\e, \bar\e)\mathcal{L} = - \frac{i}{2}\p_{\mu}[\det A (\psi \s^{\mu} \bar{\e} - \e \sigma^{\mu} \bar \psi)]\equiv \p_{\mu} {\cal J}^\mu ,
\label{Lvariation}\ee
and hence the action $S = \int d^4 x \mathcal{L}$ is invariant under those transformations.

\subsection{The VA model and the background field method}
To relate the VA model to the general background field method \cite{DeWitt:1967ub} adapted to nonlinear symmetries in \cite{Kallosh:2016qvo}, we use the follolwing dictionary: the set of all fields in the model $(\phi^i)$ includes the VA goldstino $(\psi, \bar \psi)$, and the symmetry parameters $(\xi^\alpha)$ include the constant fermions $(\e, \bar{\e})$:
\be
\f^i = (\psi^{a}, \bar \psi^{\dot{a}} )\li,\quad\xi^{\a} = (\e^{a}, \bar{\e}^{ \dot{a} })\li.
\ee
The solutions to the free field equations, special cases of $(\psi,\pb)$, are denoted by $(\la,\lb)$:
\be\label{eq.freefield}
-i(\p_\mu\la\s^\mu)_{\dot a} =0\li,\quad-i (\s^\mu \p_\mu \lb)_a =0\li.
\ee

The symmetries of the action $S(\phi)$ shown in \eqref{VALag}, in abstract form given as 
\be
\delta\phi^i= {\cal R}^i_\alpha (\phi) \xi^\alpha\li,
\label{NLS}\ee
and detailed in \rf{eq.sym12}, form the algebra
\be
\left[ {\cal R}^i_{\alpha, j} (\phi), {\cal R}^j_\beta (\phi) \right\} = f_{\alpha \beta}^\gamma {\cal R}^i_\gamma (\phi)\li.
\label{al}\ee
The functional derivative of the action has two components
\be
S_{, i} \equiv {\d S\over \d \phi^i}= \Big (\frac{ \d S }{\d \psi^{a} }, \, \frac{ \d S }{\d \bar \psi ^{\dot a} }\Big )\li,
\ee
and for the action to have a symmetry under some global transformation of the fields, the corresponding variation of the Lagrangian must be a total derivative, compare with \rf{Lvariation}:
\be
\delta \mathcal{L}={\delta S\over \delta \phi^i}{\cal R}^i_\alpha (\phi) \xi^\alpha+ \partial_\mu ({\cal J}^N)^\mu= \partial_\mu {\cal J}^\mu\li.
\ee
Here, the Noether current is
\be
({\cal J}^N)^\mu\equiv {\partial \mathcal{L}\over \partial \partial_\mu \phi^i} {\cal R}^i_\alpha (\phi) \xi^\alpha\li,
\ee
 and when the equations of motion given by the action are satisfied, $S_{,i}=0$, the current conservation follows:
\be
\partial_\mu ({\cal J}^N)^\mu- \partial_\mu {\cal J}^\mu=0\li.
\ee

The equations of motion for the background fields $(\psi, \bar \psi)$, given in \rf{eq.eq12}, each has a free part, linear in the fields, as well as a nonlinear part
\begin{subequations}\begin{align}
\frac{ \d S }{\d \psi^{a} } =- i ( \s^\mu \p_\mu \bar \psi )_{a} +i \, \tilde T^{\mu\nu} \, (\s_\nu \p_\mu \bar \psi)_{a} + i\p_{\mu} [\tilde T^{\mu\nu} (\s_\nu \bar \psi)_{a} ] \li,\\
\frac{ \d S }{\d \bar \psi^{\dot a} } =- i (\p_\mu \psi \, \s^\mu )_{\dot a} +i \, \tilde T^{\mu\nu} \, ( \p_\mu \psi \s_\nu)_{\dot a} + i\p_{\mu} [\tilde T^{\mu\nu} ( \psi \s_\nu )_{\dot a} ] \li.
\label{nonlin}\end{align}\end{subequations}
Defining the Green's function of the goldstino as the inverse of the linear term differential operator 
\be \label{eq.GreensF}
- i \s^\mu \p_\mu G (x,y)= -\delta^4(x-y)\li,
\ee
the solution for $ \psi$ is \cite{Kallosh:2016qvo}
\be\label{eq.prescribedSol}
\psi(x)= \la (x) + \int d^4 y \, G (x,y) {\delta S^\text{int}\over \d \bar \psi}\li,
\ee
where ${\delta S^\text{int}\over \d \bar \psi}$ starts with three fields. The corresponding general background field equations in DeWitt's formalism are
\be
\phi^i = \phi^i_0 + G^{ij} S_{, j }^\text{int}(\phi)\li.
\ee
Note that, ignoring derivatives, the VA theory is characterized by
\be
S^\text{int} \sim (\psi \bar \psi)^2 + (\psi \bar \psi)^3\quad\Rightarrow\quad\frac{\d S^\text{int} }{\d \psi} \sim \psi \bar \psi \bar \psi + (\psi \bar \psi)^2 \bar \psi . 
\ee
It is interesting that higher order terms with eight spinors $(\psi\bar \psi)^4$ which one would expect in the VA action \rf{VALag} are, in fact, absent. The proof of this non-trivial fact is given in Appendix A of \cite{Kuzenko:2005wh}. On the other hand, the Komargodski-Seiberg action \cite{Komargodski:2009rz}, related to the VA action \rf{VALag} by a non-linear change of variables,  does not have terms $(\psi \bar \psi)^3$ but has terms $(\psi \bar \psi)^4$, as established in \cite{Kuzenko:2010ef}.

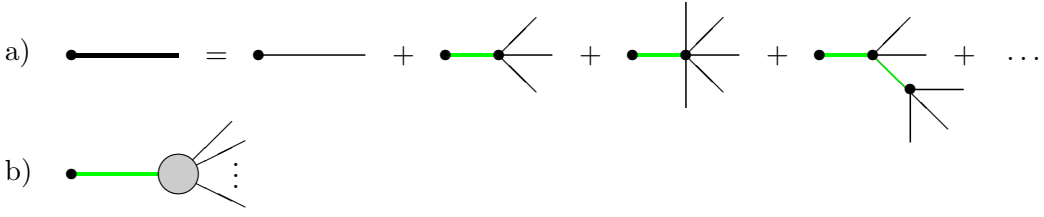
\begin{figure}[htbp]
\begin{center}
\begin{picture}(385,80)(-25,-55)
\put(-25,3){a)}
\put(0,5){\circle*{4}}\put(0,5){\linethickness{0.6mm}\line(1,0){40}}
\put(50,2.1){$=$}
\put(70,5){\circle*{4}}\put(70,5){\linethickness{0.2mm}\line(1,0){40}}
\put(120,2.1){$+$}
\put(140,5){\linethickness{0.2mm}\line(1,0){40}}\put(140,5){\linethickness{0.4mm}\color{green}\line(1,0){20}}\put(140,5){\circle*{4}}\put(160,5){\circle*{4}}\put(160,5){\linethickness{0.2mm}\line(1,1){14}}\put(160,5){\linethickness{0.2mm}\line(1,-1){14}}
\put(190,2.1){$+$}
\put(210,5){\linethickness{0.2mm}\line(1,0){40}}\put(210,5){\linethickness{0.4mm}\color{green}\line(1,0){20}}\put(210,5){\circle*{4}}\put(230,5){\circle*{4}}\put(230,5){\linethickness{0.2mm}\line(1,1){14}}\put(230,5){\linethickness{0.2mm}\line(1,-1){14}}\put(230,5){\linethickness{0.2mm}\line(0,1){20}}\put(230,5){\linethickness{0.2mm}\line(0,-1){20}}
\put(260,2.1){$+$}
\put(280,5){\linethickness{0.2mm}\line(1,0){40}}\put(280,5){\linethickness{0.4mm}\color{green}\line(1,0){20}}\put(300,5){\linethickness{0.2mm}\line(1,1){14}}\put(300,5){\linethickness{0.2mm}\line(1,-1){28}}\put(300,5){\linethickness{0.4mm}\color{green}\line(1,-1){14}}\put(280,5){\circle*{4}}\put(300,5){\circle*{4}}\put(314,-8){\circle*{4}}\put(314,-8){\linethickness{0.2mm}\line(1,0){20}}\put(314,-8){\linethickness{0.2mm}\line(0,-1){20}}
\put(330,2.1){$+$}\put(350,3){$\ldots$}
\put(-25,-42){b)}
\put(0,-40){\linethickness{0.4mm}\color{green}\line(1,0){40}}\put(0,-40){\circle*{4}}
\put(40,-40){\linethickness{0.2mm}\line(1,1){20}}\put(40,-40){\linethickness{0.2mm}\line(2,1){25}}\put(40,-40){\linethickness{0.2mm}\line(2,-1){25}}\put(60,-45){$\vdots$}
\put(40,-40){\color[rgb]{0.8,0.8,0.8}\circle*{15}}\put(40,-40){\circle{15}}
\end{picture}
\end{center}
\caption{\small Decomposition of the background field into tree diagrams (a) inspired by the depiction in \cite{Honerkamp:1971sh} of how, in an NLSM model of the VA theory, the nonlocal functional defines $\ps$ as a functional of $\la$. The figure shows the background field $\ps$ solving the nonlinear equations of motion \rf{eq.prescribedSol}. The thick line is the background field $\ps$, the thin lines are free fields $(\la,\lb)$, and the green lines are free propagators. An explicit expression for the cubic in $\la$ term with one propagator is derived in \rf{sol}. The generic interaction term is illustrated in (b).}\label{f2}
\end{figure}

Returning to the VA action, the way to solve \rf{eq.prescribedSol} is by iteration, taking into account that the solution starts with the free field, and continues by higher powers of $(\la,\lb)$, as illustrated in fig. \ref{f2}.

In the general class of models with nonlinear symmetry the following identity was established \cite{Kallosh:2016qvo} in the background field method
\be
\Big (S_{, j i_1 i_2} {\cal R}^{i_1}_\alpha {\cal R}^{i_2}_\beta+ S_{, j i_1 } {\cal R}^{i_1}_{ \gamma} f^\gamma_{\alpha \beta} +\cdots \Big )\xi^\alpha \xi^{'\beta} =0\li.
\label{doublesoft}\ee
Here, an approximation where the first term has $S_{, j i_1 i_2}$ linear in $\la$ and ${\cal R}^{i_1}_\alpha {\cal R}^{i_2}_\beta \xi^\alpha \xi^{'\beta} \sim \e'_a\bar\e_{\dot a}$ represents the double soft limit of the cubic approximation in the background field solution, and the unspecified terms vanish on shell. The identity \rf{doublesoft} therefore predicts that the double soft limit in the background field has to be described by the structure constants of the supersymmy algebra, $f^\gamma_{\alpha \beta}$. In the VA model, the corresponding algebra is the one in \rf{algebra}. We will proceed by testing this identity by finding a solution for the background field as a functional of the free field, and by studying its properties.

\section{Iterative solution for the background field in the VA model}
For convenience, we will use the expression for the VA action, quadratic and quartic in fermions, on the form given in \cite{Kuzenko:2005wh,Kuzenko:2010ef}. With $\kappa=1$ and neglecting terms containing fields to a total, higher power than four in the action, which is sufficient for an analysis of the supersymmetry variations and of the double soft limit, we have
\begin{align}
S_2&=-\frac{1}{2}\int\mathrm{d}^4\li x \langle v+\bar{v}\rangle\li,\\
S_4&=-\int\mathrm{d}^4\li x \left(\langle v\rangle\langle\bar{v}\rangle-\langle v \bar{v}\rangle\right)\li,
\end{align}
where the brackets denote trace and the derivatives only act on the nearest spinor, and
\be
{v_\mu}^\nu=i\ps\sigma^\nu\partial_\mu\pb\li,\quad \bar{v}_\mu \li^\nu=-i\partial_\mu \ps\sigma^\nu\pb\li,\\
\ee
is such that
\begin{subequations}
\begin{align}
\langle v+\bar{v}\rangle&=i(\ps\s^\mu\p_\mu\pb-\p_\mu\ps\s^\mu\pb)\li,\\
\langle v\rangle\langle\bar{v}\rangle-\langle v \bar{v}\rangle&=(\ps\s^\mu\p_\mu\pb)(\p_\nu\ps\s^\nu\pb)-(\ps\s^\nu\p_\mu\pb)(\p_\nu\ps\s^\mu\pb)\li.
\end{align}\end{subequations}
We have
\begin{align}
\frac{\delta S_2}{\delta \pb^{\dot a}}&=-\frac{i}{2}(\p_\mu\ps\s^\mu+\p_\mu\ps\s^\mu)_{\dot a}=-i(\p_\mu\ps\s^\mu)_{\dot a}\li,\label{eq.dS2}\\
\frac{\delta S_4}{\delta \pb^{\dot a}}&=2\left[(\ps\s^\nu\stackrel{\leftrightarrow}{\p}_{[\nu}\pb)(\p_{\mu]}\ps\s^\mu)_{\dot a}+(\p_{[\mu}\ps\s^\nu\p_{\nu]}\pb)(\ps\s^\mu)_{\dot a}\right]\nonumber\\
&=-(\ps\s^\nu\stackrel{\leftrightarrow}{\p}_{\mu}\pb)(\p_{\nu}\ps\s^\mu)_{\dot a}\li,
\end{align}
where, in the last step, the second variation has been simplified by a removal of terms containing the equation of motion that do not connect to the free spinor index. In \rf{eq.dS2}, with $\psi\rightarrow \la$ we get \rf{eq.freefield}, the equation of motion for the corresponding free field. The latter provides the part due to interactions
\be
\frac{\delta S^\text{int}}{\delta \pb^{\dot a}}= -(\ps\s^\nu\stackrel{\leftrightarrow}{\p}_{\mu}\pb)(\p_{\nu}\ps\s^\mu)_{\dot a} + \cdots \li ,
\label{BEOM}\ee
with the ellipses denoting terms with fields satisfying the free field equation as well as terms with five powers of the fields, originating from $S_6$.

Keeping free spinor indices implicit, letting $\bar{\sigma}$ be implied where applicable, and following \cite{Clark:1988es} we will use the identity
\be\label{eq.symmetrized}
\sigma^{(\mu}\bar{\sigma}^{\nu)}=\frac{1}{2}\left(\sigma^\mu\bar\sigma^\nu+\sigma^\nu\bar\sigma^\mu\right)=-\eta^{\mu\nu}\li.
\ee
Proceeding according to \rf{eq.prescribedSol} and acting on \rf{BEOM} with $G$, set by \rf{eq.GreensF} to be
\be
G=i\slashed{\partial}^{-1}=i\frac{\slashed{\partial}}{\p^2}\li,
\ee
we get, in the first iteration of \rf{eq.prescribedSol} where the approximation is such that each of the three spinor fields $\psi$ are taken as free fields
\begin{align}
\left.i \frac{\delta S_4}{\delta \pb}\stackrel{\leftarrow}{\slashed{\partial}}{}^{-1}\right|_{(\psi,\pb)\rightarrow(\la,\lb)}&=-\frac{i}{\p^2}\p_\delta\Big[(\la\s^\nu\stackrel{\leftrightarrow}{\p}_\mu\lb)(\p_\nu\la\s^\mu\s^\delta)\Big]\nonumber\\
&=\frac{i}{\p^2}\left[-[\p_\delta,(\la\s^\nu\stackrel{\leftrightarrow}{\p}_\mu\lb)](\p_\nu\la\s^\mu\s^\delta)+2(\la\s^\nu\stackrel{\leftrightarrow}{\p}_\mu\lb)(\p_\nu\p^\mu\la)
\right]
\end{align}
where the last term has been simplified through the addition of a term
\be
-\frac{i}{\p^2}(\la\s^\nu\stackrel{\leftrightarrow}{\p}_\mu\lb)(\p_\nu\p_\delta\la\s^\delta\s^\mu)\li,
\ee
which contains $\la\stackrel{\leftarrow}{\slashed{\p}}$, vanishing in the absence of more contractions with $\slashed{\partial}^{-1}$.

Thus, following \cite{Kallosh:2016qvo}, we find the following solution for the background field $\psi$
\be
\psi=\la+\frac{2i}{\p^2}\left[-\frac{1}{2}[\p_\delta,(\la\s^\nu\stackrel{\leftrightarrow}{\p}_\mu\lb)](\p_\nu\la\s^\mu\s^\delta)+(\la\s^\nu\stackrel{\leftrightarrow}{\p}_\mu\lb)(\p_\nu\p^\mu\la)\right]+\cdots
\label{sol}\ee
where the ellipses represent terms with fields satisfying free field equations as well as terms to higher powers than three in the free fields $(\la,\lb)$. In momentum space, with $\left(\lb(p),\la(q),\la(k)\right)$, $\partial=i\hat{p}$ and $P=(p+q+k)$, this is
\begin{align}
\ps(P) &= \la (P) +\frac{2}{P^2}\bigg[\frac{1}{2}(p+q)_\delta (p-q)_\mu k_\nu\la(k)\s^\mu\s^\delta\nonumber\\&
\hspace{2.5cm}-(p-q)_\mu k^\mu\, k_\nu\la(k)\bigg]\Big(\la(q)\s^\nu\lb(p)\Big)+\ldots
\label{solution}\end{align}

\section{Properties of the VA model background field }

\subsection{The supersymmetry rules}
A supersymmetry variation of $\psi$ in the approximation above can be studied by, as explained in general in \cite{Kallosh:2016qvo}, using $\delta \la = \epsilon$ to find $\delta \psi$. For the free fields, we find
\be
\delta \psi=\e+\frac{2i}{\p^2}\left[-\frac{1}{2}[(\e\s^\nu \p_\delta {\p}_\mu\lb)](\p_\nu\la\s^\mu\s^\delta)+(\e\s^\nu{\p}_\mu\lb)(\p_\nu\p^\mu\la)\right]+\cdots\li,
\ee
leaving the field connected to the free index intact. Here, the first term vanishes due to $\p^2\lb=0$, which follows from \rf{eq.freefield}. The second term becomes 
\be
\frac{2i}{\p^2} (\e\s^\nu{\p}_\mu\lb)(\p_\nu\p^\mu\la) =i \frac{\p^2}{\p^2} (\e\s^\nu\lb)(\p_\nu \la)= i(\e\s^\nu \lb ) \p_\nu \la\li.
\ee
If we add the corresponding for $\delta \lb= \bar{\e}$ of \rf{sol} to the above, we get a {\it local} expression for the supersymmetry transformation
\be
\d^Q(\e, \bar{\e}) \psi=\e- i( \la\s^\nu \bar {\e} - \e\s^\nu \lb ) \p_\nu \la
+\cdots\li,
\label{apprSUSY}\ee
despite that the solution for the background field is nonlocal.

Alternatively, the analysis can be performed in momentum space. Taking $q\rightarrow 0$ and $\la(q) \rightarrow \e$ in the term cubic in $\la$ in \rf{solution}, that term simplifies to
\be
\frac{2}{(p+k)^2}\Big[\frac{1}{2}p_\delta p_\mu k_\nu \la(k)\s^\mu\s^\delta - (p\cdot k) k_\nu \la(k)\Big]\Big(\e\s^\nu\lb(p)\Big)\li.
\ee
Here, with account of $p^2= k^2=0$, the first term vanishes and the second term is
\be
-\frac{2}{(p+k)^2}\left[(p\cdot k) k_\nu \la(k)\right]\left[\e\s^\nu\lb(p)\right]=-\left[\e\s^\nu\lb(p)\right] k_\nu \la(k)\li.
\ee
If we instead consider $p\rightarrow 0$ and $\lb(p) \rightarrow \e$, we get
\be
\left[\la (q)\s^\nu\bar \e \right] k_\nu \la(k)\li.
\ee
Together, these two terms reproduce \rf{apprSUSY} in momentum space.

\subsection{The double soft limit}
The double soft limit of a background field in the approximation up to cubic terms is given by \rf{solution} and
\begin{subequations}\label{eq.doublesoft}
\be
\lb(p)\rightarrow\bar{\varepsilon}\li,\quad\la(q)\rightarrow\varepsilon'\li,\quad p,q\rightarrow0\li:
\ee
\be
\ps(k) = \la(k)-\frac{k\cdot(p-q)}{k\cdot(p+q)}(\varepsilon'\s^\nu\bar{\varepsilon})k_\nu\la(k) \li.
\ee
\end{subequations}
The first term in the square brackets in \rf{solution} has an extra softness compared to the second one as it depends on $(p+q)_\delta (p-q)_\mu$ rather than $(p-q)_\mu$ alone. The second term, as predicted by the background field method in \cite{Kallosh:2016qvo}, shows up as a consequence of the algebra given in \rf{algebra}. Thus, we have shown that the identity \rf{doublesoft}, defining the double soft limit of the background field, is valid in the VA model to lowest order in the free fields.

\section{Discussion}
There is an increasing interest in nonlinear supersymmetries since they are helpful in building cosmological models, but also since the LHC at present has not yet discovered the superpartners of the known particles. The first model of nonlinear supersymmetries were discovered by Volkov and Akulov at about the same time as models with linear supersymmetry. However, the nonlinear models were not studied as much as the linear ones. 

In this paper we have made a step towards such a study by detailing in the example of the VA model the general background field method developed in \cite{Kallosh:2016qvo} for nonlinear symmetries. We have solved the equation of motion for the background field up to cubic approximation in the free fields, and we have studied the solution, finding it consistent with the expected nonlinear supersymmetry and that the variation of the nonlocal background field in our approximation produces the expected local nonlinear supersymmetry transformations. We have also studied the double soft limit of the background field and found an agreement with the prediction for a double soft limit following from general identities derived in \cite{Kallosh:2016qvo}. More studies of this kind will be possible in the future. Of relevance is, for example, to study different models, and the multi soft behaviour of the amplitudes. A more important venue of research is to develop the background field techniques for models with nonlinear symmetries beyond tree level.

\acknowledgments{We thank  P.~Binetruy, B.~de Wit, Y.-t.~Huang, R.~Roiban, B.~Vercnocke and T.~Wrase for useful discussions, as well as the participants of the workshop `Supergravity: what next?' at the Galileo Galilei Institute for Theoretical Physics (GGI). This work is supported by the SITP and by the NSF Grant PHY-1316699. AK is also supported by the K. A. Wallenberg Foundation. We also thank the GGI for hospitality and the INFN for partial support during the completion of this work.} 

\bibliographystyle{JHEP}
\bibliography{refs}
\end{document}